\documentclass[letterpaper,floatfix,aps,prb,amsmath,amssymb,twocolumn,superscriptaddress,nopacs]{revtex4-2}

\usepackage{graphicx}
\usepackage{verbatim}
\usepackage{color}
\usepackage{array}
\usepackage[pdfauthor={Catelani,Glazman,Schoelkopf},
            pdftitle={Ac losses in field-cooled type I superconducting cavities}]{hyperref}

\hypersetup{pdfstartview={XYZ null null 1.05}}

\newcommand{\be}{\begin{equation}}
\newcommand{\ee}{\end{equation}}
\newcommand{\bea}{\begin{eqnarray}}
\newcommand{\eea}{\end{eqnarray}}

\begin{document}

\title{Ac losses in field-cooled type I superconducting cavities}

\author{G. Catelani}
\affiliation{JARA Institute for Quantum Information (PGI-11), Forschungszentrum J\"ulich, 52425 J\"ulich, Germany}
\affiliation{Yale Quantum Institute, Yale University, New Haven, Connecticut 06520, USA}

\author{K. Li}
\affiliation{Departments of Physics and Applied Physics, Yale University, New Haven, Connecticut 06520, USA}

\author{C. J. Axline}
\altaffiliation[Present address: ]{Institute for Quantum Electronics,
ETH Z\"urich, Otto-Stern-Weg 1, 8093 Z\"urich, Switzerland}
\affiliation{Departments of Physics and Applied Physics, Yale University, New Haven, Connecticut 06520, USA}

\author{T. Brecht}
\altaffiliation[Present address: ]{HRL Laboratories, LLC, 3011 Malibu Canyon Road, Malibu, California 90265, USA}
\affiliation{Departments of Physics and Applied Physics, Yale University, New Haven, Connecticut 06520, USA}

\author{L. Frunzio}
\affiliation{Departments of Physics and Applied Physics, Yale University, New Haven, Connecticut 06520, USA}

\author{R. J. Schoelkopf}
\affiliation{Departments of Physics and Applied Physics, Yale University, New Haven, Connecticut 06520, USA}

\author{L. I. Glazman}
\affiliation{Departments of Physics and Applied Physics, Yale University, New Haven, Connecticut 06520, USA}

\begin{abstract}
As superconductors are cooled below their critical temperature, stray magnetic flux can become trapped in regions that remain normal. The presence of trapped flux facilitates dissipation of ac current in a superconductor, leading to losses in superconducting elements of microwave devices. In type II superconductors, dissipation is well-understood in terms of the dynamics of vortices hosting a single flux quantum. In contrast, the ac response of type I superconductors with trapped flux has not received much attention. Building on Andreev's early work~\cite{Andreev67}, here we show theoretically that the dominant dissipation mechanism is the absorption of the ac field at the exposed surfaces of the normal regions, while the deformation of the superconducting/normal interfaces is unimportant. 
We use the developed theory to estimate the degradation of the quality factors in field-cooled cavities, and we satisfactorily compare these theoretical estimates to the measured field dependence of the quality factors of two aluminum cavities.
\end{abstract}

\date{\today}

\maketitle

\section{Introduction}

Superconducting cavities are under intense investigation for diverse applications such as particle accelerators~\cite{Padamsee} and quantum information processing~\cite{Ibarcq}. A fundamental question of practical importance is what limits their quality factors, or equivalently what mechanisms are responsible for power dissipation. In type II superconductors, it has long been recognized that one such mechanism is the motion of vortices, also known as flux flow. Vortices are generically present when a superconductor is cooled in a magnetic field $B_0$. The corresponding dissipative losses can be characterized in terms of the flow resistivity $\rho_f$. Following empirical suggestions, $\rho_f$ is expected to be proportional to the normal-state resistivity $\rho_n$ and the ratio between magnetic field $B_0$ and second critical field $B_{c_2}$,
\be\label{eq:rhof}
\rho_f = \rho_n \frac{B_0}{B_{c_2}}\,.
\ee
Theoretical justification of Eq.~(\ref{eq:rhof}), attributing $\rho_f$ to the losses in the normal cores of moving vortices was given by Bardeen and Stephen, see \cite{BarSteph1965} and references therein. For high-field applications, such as superconducting magnets and RF cavities for particle accelerators, the flux-flow dissipation can significantly impact performance. By pinning vortices, the dissipation can be reduced, and this has led to intense and still ongoing research into ways to pin vortices~\cite{SuSTFocusAPC}, and into surface treatments~\cite{Kelly2017} to remove defects facilitating vortex penetration into the bulk of superconducting cavities~\cite{Liarte2017}.

In the case of superconducting devices for quantum information processing the fields involved are usually small, as external magnetic fields are carefully screened. The most widely used material for superconducting quantum devices -- including bulk cavities -- is aluminum, which is a type I superconductor. But when cooled in a field, type I superconductors can enter into the so-called intermediate state, in which normal-state regions are interspersed among superconducting ones, allowing the passage of magnetic flux. These normal regions can take different shapes, such as flux tubes, resembling vortices, or laminar domains; a number of theoretical and experimental studies have focused on this aspect of the intermediate state over the years~\cite{Landau37,Dorsey98,Prozorov2007}, and on dc losses~\cite{Solomon69,Lerski75}. Surprisingly little attention has been given to the question of ac losses in the intermediate state 
associated with these normal regions, with the notable exception of the seminal works by Andreev and collaborators~\cite{Andreev67,Andreev68,Andreev71}. Here we revisit this issue to give a unified picture of ac dissipation in superconductors, and to compare the theory to the results of experiments performed with type I cavities.

In the next section we briefly review the model of Ref.~\cite{Gurevich} for ac losses by flux flow in type II superconductors.
In Sec.~\ref{sec:typeI} we summarize Andreev's result for dissipation in type I superconductors~\cite{Andreev67} and contrast it to that in type II materials. In Sec.~\ref{sec:Qfactor} we first derive an estimate for the dependence of a superconducting cavity quality factor $Q$ on cooling field, which is then compared to experiments with bulk aluminum cavities of two different geometries. We summarize and discuss our findings in Sec.~\ref{sec:summary}.

\section{ac dissipation in type II superconductors}
\label{sec:tpyeII}

As mentioned in the Introduction, in type II superconductors the pinning of vortices has long been investigated as a way to reduce flux-flow dissipation. Early models treated pinning by introducing, for example, a harmonic confining potential~\cite{GittRose66}. In such a model, there is a depinning frequency 
above which
pinning is ineffective and the dissipated power saturates to a frequency-independent value. Below the depinning frequency, the ac dissipated power increases quadratically with the frequency $\omega$.
More recently, a model
accounting for strong pinning centers predicted not only a quadratic increase with frequency at the lowest frequencies and a saturation at high frequencies, but also an intermediate frequency domain with a $\sqrt{\omega}$ dependence for the dissipated power. Here for simplicity we neglect the effect of pinning~\cite{note1}, thus dispensing with the low-frequency quadratic asymptote for the dissipated power, and refer the reader to Ref.~\cite{Gurevich} for its treatment. The extension to random, weak pinning centers of different dimensionalities can be found in Ref.~\cite{Liarte2018}. We also disregard the exponentially small effect of quasiparticles which are assumed to be at thermal equilibrium with temperature much smaller than the superconducting gap divided by the Boltzmann constant, $T\ll\Delta/k_B$; thus the ac dissipation is dominated by the flux-flow contribution. This contribution is a potential explanation for the observation that in many experiments with thin-film quantum devices, the internal dissipation is no longer exponentially improving below $\sim 100-200\,$mK~\cite{note2}.

The model we consider accounts for the viscous motion of a vortex line under the action of the ac field; denoting by $u(z,t)$ the displacement of an infinitesimal vortex element 
at depth $z$ from its position in the absence of the ac field, the equation of motion for $u$ reads
\begin{equation}\label{eq:uGur}
  \eta \dot{u} = \varepsilon u'' + F e^{-z/\lambda}e^{-i\omega t}\, .
\end{equation}
Here $\dot{u}$ and $u''$ are, respectively, the time derivative and the second spatial derivative of the displacement,
$\lambda$ is the penetration depth, $\eta$ is the drag coefficient, $\varepsilon$ is the vortex line tension, and $F$ is 
the Lorentz force (per unit length of
the vortex line) at $z=0$; this force acting on the vortex line is due to the ac field parallel to the superconductor's surface. The textbook expressions for these last three quantities 
can be found, {\sl e.g.}, in 
Ref.~\cite{Tinkham}. Note that $u$ is in general a two-dimensional vector, but here we treat it as a scalar in the direction of the applied force, a direction which is assumed constant in time.

The drag coefficient $\eta$ originates from dissipation in the normal core of the vortex. The dissipation is caused by the electric field generated by the vortex motion (Sec.~5.5.1 in~\cite{Tinkham}),
\begin{equation}\label{eq:vdrag}
    \eta = \frac{\Phi_0^2}{2\pi \xi^2 \rho_n}\, .
\end{equation}
Here $\rho_n$ is the normal-state resistivity and $\xi$ is the coherence length, which gives approximately the radius of the normal core. The line tension
\begin{equation}\label{eq:tensionII}
    \varepsilon = \frac{\Phi_0^2}{4\pi \mu_0 \lambda^2} \ln \kappa
\end{equation}
is given by the free energy per unit length of a static vortex; the term proportional to it in Eq.~(\ref{eq:uGur}) accounts for the energy cost of elastic deformation of the vortex. The line tension is mainly due to the kinetic energy of the superfluid current around the vortex, see Sec.~5.1.2 in~\cite{Tinkham}. There is an additional contribution to $\varepsilon$ from the vortex core, which is neglected there; we show explicitly in Appendix~\ref{app:coretension} that this is a good approximation in strongly type II superconductors with Ginzburg-Landau parameter $\kappa =\lambda/\xi \gg 1$. Finally,
the magnitude $F$ of the Lorentz force acting on the vortex line is proportional to the magnitude $H_p$ of the parallel to the surface alternating magnetic field of the impinging electromagnetic wave
(see Sec.~5.2 in~\cite{Tinkham}),
\begin{equation}\label{eq:vforce}
    F = \Phi_0 H_p/\lambda \,.
\end{equation}

We are interested in calculating the power $P_v$ dissipated by a vortex as it moves and bends under the action of the ac field. To this end, we need to integrate over $z$ the product $Fe^{-z/\lambda}\dot{u}$ of the force and velocity. Therefore, we first solve Eq.~(\ref{eq:uGur}) for $u$ by performing a Fourier transform,
\begin{equation}
  u(z,t) = e^{-i\omega t}\int \frac{dk}{2\pi} \, \tilde{u}(k) e^{ikz}\, ,
\end{equation}
with the boundary condition $u'(0,t)=0$ corresponding to no surface pinning. We do not present here explicitly the mathematical derivation of the final expression for $P_v$, as it is a simplified version of that given in Ref.~\cite{Gurevich}. We note, however, that the line tension $\varepsilon$ in general depends on $k$, but that this dependence can be neglected at low frequency, such that $\omega \lambda^2 \eta/\varepsilon \ll 1$.
In this regime, the dissipated power $P_v$ is
\begin{equation}
    P_v = \frac12 (\lambda F)^2 \sqrt{\frac{\omega}{2\eta\varepsilon}}
\end{equation}
which, using Eqs.~(\ref{eq:vdrag})-(\ref{eq:vforce}), agrees with the corresponding result in~\cite{Gurevich}. We mention in passing that the $\sqrt\omega$ frequency dependence of the dissipated power is not unique to the vortex flow mechanism; on its own, it is insufficient to distinguish it, \textit{e.g.}, from the quasiparticle losses at low (effective) temperature $k_BT \ll \omega$.

It is important to note that even at finite frequency $\omega$ the dissipation in a type II material is associated with the motion of vortices, rather than with the penetration of the impinging electromagnetic wave into the normal core of a static vortex. The reason is that the skin depth (which is in general longer than the penetration depth in the superconducting state) exceeds greatly the core radius $\xi$, which makes such penetration impossible.

In the following, we will compare the ac dissipation in type I and type II superconductors, which we denote as $\tilde{P}_\mathrm{I}$ and $\tilde{P}_\mathrm{II}$, respectively. For such a comparison, we define $\tilde{P}_\mathrm{I}$ and $\tilde{P}_\mathrm{II}$ as the dissipated power per unit surface area of a superconductor cooled in a field $B_0$. Considering a sample of area $S$, the number of vortices $N_v$ is given by the ratio of flux to the flux quantum,
\begin{equation}\label{eq:Nv}
    N_v = \frac{B_0 S}{\Phi_0}\, .
\end{equation}
Then the dissipated power per unit area $\tilde{P}_\mathrm{II}$ is
\begin{equation}\label{eq:PSII}
   \tilde{P}_\mathrm{II} = \frac{P_v N_v}{S} = \frac{B_0}{B_c} \frac{H_p^2}{2} \sqrt{\frac{\mu_0\omega\rho_n}{\ln\kappa}} \, ,
\end{equation}
where
\begin{equation}
  B_c = \frac{\Phi_0}{2\pi\xi\lambda}
\end{equation}
is the thermodynamic critical field. In the next section we show that a similar expression holds for $\tilde{P}_\mathrm{I}$.

\section{ac dissipation in type I superconductors}
\label{sec:typeI}

In contrast to the vortices with the radius of the order of coherence length $\xi$, normal regions in type I superconductors can be of a macroscopic size. As already shown in the early work by Landau~\cite{Landau37}, for the laminar configuration the width $w_n$ of the normal parts is of the order of $w_n \sim \sqrt{d\delta}$, with $\delta\approx \xi -\lambda$ and $d$  being the sample thickness (see Sec.~2.3.2 in~\cite{Tinkham}); for a strongly type I superconductor, $\kappa \ll 1$, we have $\delta \sim \xi$. Neglecting again pinning, the normal regions can move and, as shown in Ref.~\cite{Andreev67}, due to this motion there is flux-flow dissipation in the presence of a dc current, and the dc resistance in the intermediate state is proportional to the normal-state fraction $x_n=B_0/B_c$ of the sample. This mechanism of the dc dissipation, which was later extended to flux tubes~\cite{Andreev71}, is no different than the one at work in type II superconductors. 
Interestingly, at a sufficiently high frequency the ac dissipation in a type I superconductor does not involve motion of the normal regions. Once the normal-state skin depth becomes shorter than $w_n$, the impinging field penetrates into the static normal domains thus producing the dissipation. 

The surface resistance $R_s$ is associated with the skin effect in the normal domains~\cite{Andreev67},
\begin{equation}\label{eq:Rs}
R_s = \sqrt{\frac{\mu_0\omega \rho_n}{2}} \frac{B_0}{B_c} \, ,
\end{equation}
where the square root factor is the normal-state surface resistance~\cite{note3} and $B_0/B_c$ is the normal fraction $x_n$. The power dissipated per unit area $\tilde{P}_\mathrm{I}$ for a type I superconductor is then
\begin{equation}\label{eq:PSI}
\tilde{P}_\mathrm{I} = \frac12 R_s H_p^2 = \frac{B_0}{B_c} \frac{H_p^2 }{2} \sqrt{\frac{\mu_0\omega\rho_n}{2}} \, .
\end{equation}
Note the similarity between Eqs.~(\ref{eq:PSII}) and (\ref{eq:PSI}), despite the fact that the dissipation mechanisms are quite different: in type II superconductors the power is determined by the interplay between the elastic deformation of the vortex core and the ohmic loss in it, while in type I superconductors the power is given by the local loss at the surface of the normal-state regions exposed to the ac field.

It was claimed~\cite{Andreev67} that the deformation of the interface between normal and superconducting region does not significantly contribute to losses in type I superconductors; we show next that this is indeed the case.

\subsection{Deformation of the normal/superconductor interface}
\label{sec:t1def}

Let us consider a planar domain wall separating a normal region from a superconducting one; in the intermediate state, the magnetic field is zero in the superconducting part and $B_c$ in the normal one. The equation governing the displacement $u(z,t)$ of the wall has the same form as for vortices,
\begin{equation}\label{eq:bending}
    \tilde{\eta} \dot{u} = \tilde{\varepsilon} u'' + \tilde{F} e^{-z/\lambda} e^{-i\omega t}
\end{equation}
where the parameters with tilde are per unit area instead of per unit length as in Eq.~(\ref{eq:uGur}). (The displacement $u$ is in principle also a function of the other direction, $x$, along the wall, but since the force depends only on $z$, there is translational invariance along $x$.) Similar to the vortex case of the previous section, the dissipated power $P_w$ (per unit length along the wall) associated with the deformation of the domain wall is
\begin{equation}\label{eq:power}
    P_w = \frac12 (\lambda \tilde{F})^2 \sqrt{\frac{\omega}{2\tilde{\eta}\tilde{\varepsilon}}} \, .
\end{equation}
Next we address the question of how to estimate the parameters in the right hand side of Eq.~(\ref{eq:power}) for a type I superconductor.

As with any domain wall separating two phases, there is a surface energy $\gamma$ associated with the domain wall. In the present case the surface energy is (see  also Sec.~4.3 in~\cite{Tinkham})
\begin{equation}\label{eq:surfen}
    \gamma = \delta \frac{B_c^2}{2\mu_0}\, ,
\end{equation}
with $\delta \approx \xi -\lambda$, as introduced at the beginning of this section. Therefore we estimate the surface tension for a domain wall in a type I superconductor as
\begin{equation}\label{eq:tension}
    \tilde{\varepsilon} = \xi \frac{B_c^2}{2\mu_0}\, .
\end{equation}

\begin{figure}
    \centering
    \includegraphics[width=0.44\textwidth]{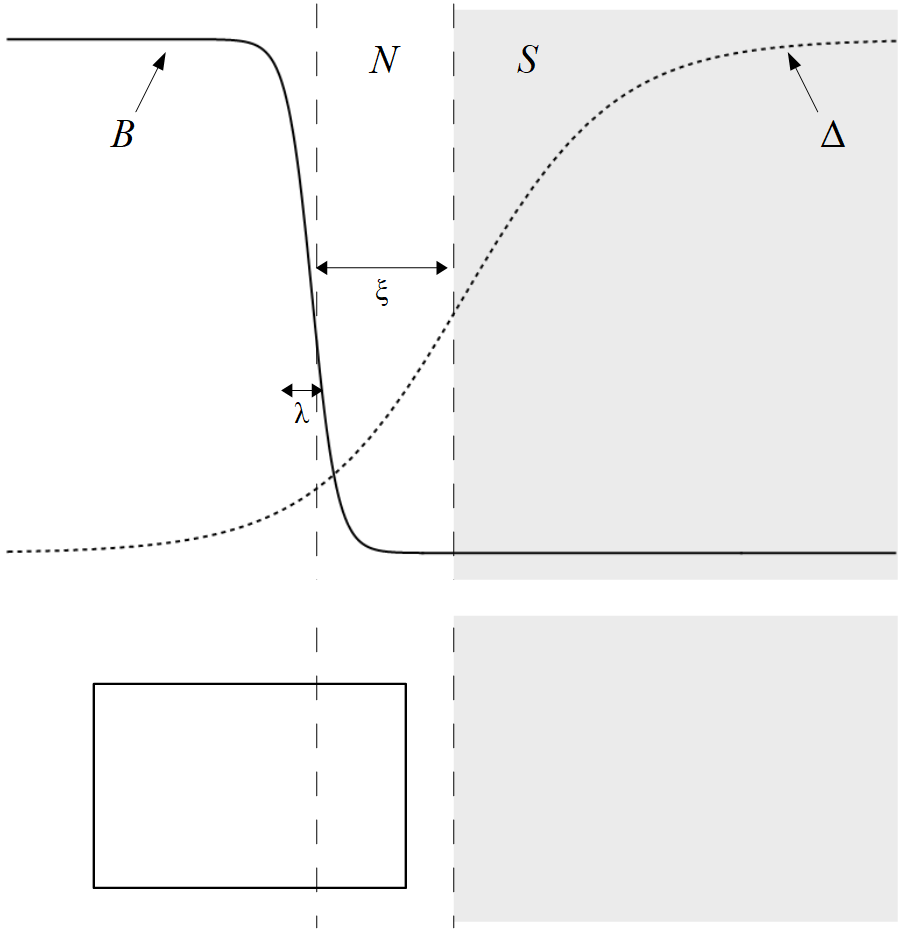}
    \caption{Top: schematic representation of a domain wall. From left to right, the magnetic field $B$ (solid line) decreases from $B_c$ to zero over the penetration depth $\lambda$, while the order parameter $\Delta$ (dotted line) rises from zero to a finite value over the much longer coherence length $\xi$. The superconducting region $S$ is shaded, while the normal region $N$ with finite electric field is comprised between the two vertical dashed lines. Bottom: the rectangular integration contour used to calculate the electric field, as viewed from above the surface of the superconductor. Contours fully to the left (right) of the left (right) dashed line give no electric field in the regions outside the two vertical lines.}
    \label{fig:dw}
\end{figure}

For the drag coefficient $\tilde{\eta}$, we can reason in a way similar to that for vortices (Sec.~5.5.1 in~\cite{Tinkham}), but adapting it to the domain wall. As for vortices, we attribute the drag to Joule heating in the normal part, and if the wall is moving with speed $v_D$ in the direction normal to the wall, the dissipated power per area is then $\tilde{\eta} v_D^2$. Here the relevant normal part is the region of thickness $\xi$ over which the order parameter is rising, while the magnetic field has already decreased significantly over the shorter length $\lambda$, see Fig.~\ref{fig:dw}.
Indeed, if we take an arbitrary contour fully contained in the normal region with magnetic field $B_c$, the flux through the enclosed area does not change in time, so there is no electric field in that region. On the other hand, by considering a rectangle with a side in the normal region (of thickness $\xi$)  with field $B_c$ and the parallel side in the normal region with no field, by Maxwell equations the electric field $E$ in the latter region has magnitude
\begin{equation}
    E = B_c v_D \, .
    \label{eq:E}
\end{equation}
According to Ohm's law, the dissipated power density is $E^2/\rho_n$. Integrating it over the thickness $\xi$ of the normal region with electric field $E$ of Eq.~(\ref{eq:E}) and equating the result to $\tilde{\eta} v_D^2$, we arrive at the estimate
\begin{equation}\label{eq:drag}
    \tilde{\eta} = \xi \frac{B_c^2}{\rho_n} \, .
\end{equation}

We estimate the force per unit area $\tilde{F}$ acting on the domain wall by evaluating the difference between the energy density $\mathcal{E}$ in the presence of a small external quasi-static field of magnitude $H_p \ll B_c/\mu_0$ and the energy density in its absence [cf. Eq.~(6) in~\cite{Andreev67}]. The energy density is given by the kinetic energy density $\mu_0\lambda^2 j^2/2$ of the current density $j$ in the domain wall. Without the ac field, we have the field going from $B_c$ to zero over distance $\lambda$ (see Fig.~\ref{fig:dw}); then by Maxwell equation, the magnitude of the screening current in that region is $j_s\approx B_c/\mu_0\lambda$. Once the quasi-static ac field is applied, there is additional current $j_p \approx (H_p/\lambda) e^{-z/\lambda}e^{-i\omega t}$. Assuming that the ac field is perpendicular to the wall, the additional current is parallel to $j_s$, and the change in energy density is
\begin{align}
    \Delta \mathcal{E} & = \frac12 \mu_0 \lambda^2\left[(j_s + j_p)^2 - \lambda^2 j_s^2\right] \approx \mu_0 \lambda^2 j_s j_p \nonumber \\
    & \approx B_c H_p e^{-z/\lambda}e^{-i\omega t} \, .
\end{align}
Therefore the magnitude of the force acting on the domain wall is
\begin{equation}\label{eq:force}
    \tilde{F} = B_c H_p \, .
\end{equation}
Note that if $H_p$ has a component $H_{pp}$ parallel to the wall, the added current is perpendicular to the wall, and its contribution to the energy density is smaller by a factor $\mu_0 H_{pp}/B_c \ll 1$, so we neglect its effect.

Interestingly, the force per unit area $\tilde{F}$ could be simply obtained by dividing the force per unit length on a vortex, Eq.~(\ref{eq:vforce}), by the circumference of the vortex core, $\tilde{F} = F/2\pi\xi$. For the drag and surface tension, on the other hand, there are additional dependencies on $\kappa$: comparing Eq.~(\ref{eq:vdrag}) to Eq.~(\ref{eq:drag}) and Eq.~(\ref{eq:tensionII}) to Eq.~(\ref{eq:tension}) we find $\tilde{\eta} = \eta/2\pi\xi\kappa^2$ and $\tilde{\varepsilon} = \varepsilon/2\pi\xi \ln \kappa$.

We have now all the ingredients needed to estimate the dissipated power per unit area associated with the deformation of the interfaces between normal and superconducting domains. For each superconducting domain, there are two such interfaces, and the number of domains per unit length (\textit{i.e.}, their linear density) can be written as $B_0/B_c w_n$, accounting for the ``squeezing'' of the cooling field $B_0$ into regions of width $w_n$ where the field reaches the critical value $B_c$ (see Sec.~2.3.3 in~\cite{Tinkham}). Therefore, the dissipated power per unit area $\tilde{P}_w$ originating from the domain walls is
\begin{equation}
  \tilde{P}_w = 2P_w \frac{B_0}{B_c w_n} = \tilde{P}_\mathrm{I} \, 2\kappa^2 \frac{\sqrt{2}\xi}{w_n}\, ,
\end{equation}
where we have substituted Eqs.~(\ref{eq:tension}), (\ref{eq:drag}), and (\ref{eq:force}) into Eq.~(\ref{eq:power}), and used Eq.~(\ref{eq:PSI}).
For a type I superconductor,  $2\kappa^2 < 1$ and if the laminae are macroscopic, $w_n \sim \sqrt{d \xi}$ with $d \gg \xi$ being the sample thickness, then $\xi/w_n \ll 1$.
Therefore we expect that $\tilde{P}_{w} \ll \tilde{P}_\mathrm{I}$: the power loss from deformation of the interfaces can be neglected, which confirms Andreev's assertion~\cite{Andreev67}.

The above result holds also if the normal regions are flux tubes rather than laminae. In this case, the parameters (per unit length) entering Eq.~(\ref{eq:uGur}) can be approximately obtained by multiplying those (per unit area) in Eqs.~(\ref{eq:tension}), (\ref{eq:drag}), and (\ref{eq:force}) by $\pi R_t$, with $R_t$ the tubes' radius (see also Appendix~\ref{app:coretension}); this leads to a dissipated power $\pi R_t P_w$ per flux tube. Since the number of tubes per unit area is $B_0/B_c \pi R_t^2$, we find
\begin{equation}
   \tilde{P}_t = \tilde{P}_\mathrm{I}\, 2\kappa^2 \frac{\sqrt{2}\xi}{R_t}
\end{equation}
for the dissipated power per unit area from the deformation of the interfaces of the tubes; this power $\tilde{P}_t$
can again be neglected in comparison to $\tilde{P}_\mathrm{I}$.

\subsection{Dissipation as function of \texorpdfstring{$\kappa$}{k}}
\label{sec:Pkappa}

As already remarked after Eq.~(\ref{eq:PSI}), the main contributions $\tilde{P}_\mathrm{I}$ and $\tilde{P}_\mathrm{II}$ to the dissipated power in the presence of trapped flux are similar for type I and type II superconductors, although the mechanisms determining them are different. In fact, let us introduce the powers per unit area $\tilde{P}_d$ arising from the deformation of the normal/superconductor interface and $\tilde{P}_s$ due to direct, local loss from the parts of the surface which are in the normal state. So far we have seen that
\begin{equation}
\tilde{P}_d \approx \left\{
\begin{array}{ll}
  \tilde{P}_\mathrm{I} \, 2\kappa^2 \sqrt{2}\xi/R_t\, , & \kappa \ll 1 \\
  \tilde{P}_\mathrm{I} \sqrt{2/\ln \kappa}\, , & \kappa \gg 1
\end{array}
\right.
\end{equation}
where in the top line we consider for concreteness flux tubes, and the bottom line follows from Eqs.~(\ref{eq:PSII}) and (\ref{eq:PSI}).
Similarly, we can write
\begin{equation}
\tilde{P}_s \approx \left\{
\begin{array}{ll}
  \tilde{P}_\mathrm{I} \, , & \kappa \ll 1 \\
  \tilde{P}_\mathrm{I}/2\kappa \, , & \kappa \gg 1
\end{array}
\right.
\end{equation}
where the bottom line is obtained by estimating the normal-state fraction in a sample of area $S$ as $N_v \pi \xi^2/S=B_0/2\kappa B_c$, with $N_v$ of Eq.~(\ref{eq:Nv}), and using this estimate (instead of $x_n = B_0/B_c$) in Eqs.~(\ref{eq:Rs}) and (\ref{eq:PSI}). Note that we are here assuming that penetration of the impinging wave into the vortex core is possible, which likely overestimates $\tilde{P}_s$ for $\kappa \gg 1$, cf. Sec.~\ref{sec:tpyeII}. 

As discussed above, we have $\tilde{P}_s \gg \tilde{P}_d$ for $\kappa \ll 1$, while we find $\tilde{P}_s \ll \tilde{P}_d$ for $\kappa \gg 1$.
Interestingly, at $\kappa$ of order unity both mechanisms give contributions of similar order. In the case of the power $\tilde{P}_d$ for flux tubes in the type I regime, this can be seen as follows: for small $\kappa$ $R_t$ is macroscopic, $R_t \sim \sqrt{d\delta} \approx \sqrt{d\xi}$ (we remind that $d$ is the sample thickness and $\delta \approx \xi-\lambda$); as $\kappa$ increases, the coherence length $\xi$ and the penetration depth $\lambda$ become of the same order, and therefore $\delta \to 0$. However, the coherence length is the minimum length over which the order parameter can vary, so we must always have $R_t \gtrsim \xi$, and the inequality will saturate for values of $\kappa$ of order unity.
In summary, for the total dissipated power per unit area $\tilde{P} = \tilde{P}_d+\tilde{P}_s$ we find a weak dependence on $\kappa$, since we have shown that there is a smooth crossover from $\tilde{P}_\mathrm{I}$ at small $\kappa$ to $\tilde{P}_\mathrm{II} = \tilde{P}_\mathrm{I}\sqrt{2/\ln\kappa}$ at large $\kappa$. In the next section we consider how this dissipated power affects the quality factor of a superconducting cavity.

\section{Dependence of cavity quality factor on cooling field}
\label{sec:Qfactor}

The quality factor $Q$ of a resonant systems is defined as the ratio between the energy $U$ stored in the resonator over the energy loss per unit cycle $P_\mathrm{tot}/\omega$,
\begin{equation}
Q = \frac{U\omega}{P_\mathrm{tot}} \, ,
\end{equation}
where $P_\mathrm{tot}$ and $\omega$ are the total dissipated power and the angular frequency, respectively. As discussed in the Introduction, reaching a high quality factor is useful in many applications. Here we focus on the contribution $P$  to the dissipated power originating from trapped flux, $P_\mathrm{tot} = P_0 + P$, where $P_0$ denotes the power loss in the absence of trapped flux (due, for example, to dielectric losses, two-level systems, etc.). We have seen in the previous sections that $P$ is proportional to the cooling field $B_0$; therefore we can separate the inverse quality factor of a superconducting cavity into a zero-field part and a field-dependent part:
\be\label{eq:Qi}
\frac{1}{Q} = \frac{1}{Q_0} + \frac{1}{Q(B_0)}
\ee
with
\be\label{eq:QiB0}
\frac{1}{Q(B_0)} = \frac{P}{\omega U} \equiv \alpha B_0\,.
\ee
Here we have introduced the coefficient $\alpha$, which measures how the quality factor degrades as the cooling field increases. The dissipated power $P$ is obtained by integrating the dissipated power per unit area $\tilde{P}$ over the internal cavity surface $S$, while the stored energy $U$ can be calculated as the magnetic energy in the volume $V$ enclosed in the cavity; therefore $\alpha$ is given by
\begin{equation}\label{eq:a_def}
\alpha = \frac{\int_S d^2r\, \tilde{P}}{\omega B_0 \frac{\mu_0}{2}\int_V d^3r\, H^2(r)  }
\end{equation}
with $H$ being the magnetic field inside the cavity. We focus henceforth on type I superconductors, $\tilde{P} \approx \tilde{P}_\mathrm{I}$, to facilitate comparison with experimental data; the corresponding results for $\alpha$ at large $\kappa$ can be obtained by multiplying our findings by $\sqrt{2/\ln\kappa}$, see Sec.~\ref{sec:Pkappa}.

We can separate $\alpha$ into a material-dependent factor and a geometry-dependent one: substituting Eq.~(\ref{eq:PSI}) into Eq.~(\ref{eq:a_def}), we write the result as
\begin{equation}\label{eq:a_factors}
  \alpha = \frac{\delta_s(\omega)}{2B_c} G
\end{equation}
where
\begin{equation}\label{eq:skindepth}
  \delta_s (\omega) = \sqrt{\frac{2\rho_n}{\mu_0 \omega}}
\end{equation}
is the skin depth and 
\begin{equation}\label{eq:Gdef}
G = \frac{\int_{S_f} d^2r \, H_p^2(r)}{\int_V d^3r\, H^2(r) }
\end{equation}
has units of inverse length~\cite{note4}. Note that to the integral in the numerator contribute only those parts of the internal cavity surface, denoted with $S_f$, where there is trapped flux passing through the surface; then surfaces parallel to the cooling field are in general excluded from the integral. Moreover, there has to be a finite parallel component $H_p(r)$ of the magnetic field at the active surfaces.

The appearance of the skin depth $\delta_s$ in Eq.~(\ref{eq:a_factors}) can of course be traced back to the fact that for type I superconductors the power loss is dominated by the normal-state parts of the surface, see Eq.~(\ref{eq:Rs}). The formula (\ref{eq:skindepth}) for $\delta_s$ is valid for the normal skin effect, when $\delta_s \gg \ell$, where $\ell$ is the mean free path. If this inequality is not satisfied, one should use instead the formula for the \textit{anomalous} skin effect~\cite{LL10}. Up to a numerical factor $\sim 1$, the anomalous skin depth is
\begin{equation}\label{eq:anom_skin}
  \delta_{s,a} (\omega) = \left[\delta_{s}^2(\omega) \ell \right]^{1/3}\,;
\end{equation}
it is independent of the mean free path, since $\rho_n \ell = 3/\nu e^2 v_F$ with $\nu$ and $v_F$ being the density of states at the Fermi energy and the Fermi velocity, respectively.

\subsection{Measurements of quality factor vs cooling field}

\begin{figure}
  \centering
  \includegraphics[width=0.47\textwidth]{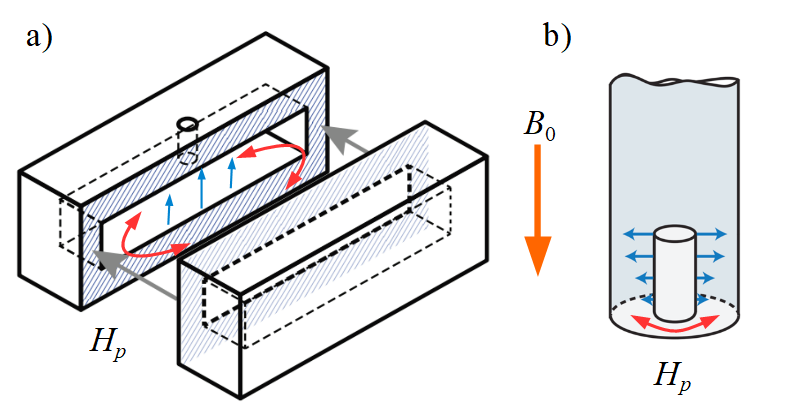}
  \caption{Schematic representation of the a) rectangular and b) coaxial cavities. The vertical orange arrows gives the direction of the cooling field $B_0$. The straight blue arrows represent the electric field of the measured mode and the red curved ones the magnetic field $H_p$.}\label{fig:cavities}
\end{figure}
 
To quantitatively test the theory described above, we outline here measurements of the quality factor of two cavities cooled in the presence of a magnetic field $B_0$, as reported in Ref.~\cite{MM18}. Two cavities of different shapes were fabricated by machining holes into blocks of high purity (4N) aluminum.  One cavity is rectangular in shape, see diagram in Fig.~\ref{fig:cavities}a, and consists of two halves joined by a seam. The cooling field was applied parallel to the seam, so that normal domains can form without crossing it; this was done to minimize possible field-dependent losses at the seam. For the same reason, the quality factor of the TE$_{101}$ mode with frequency $\sim 9.7\,$GHz was measured; the measurement technique was described in Ref.~\cite{reagor}.
The second cavity was a $\lambda/4$ coaxial cavity,  Fig.~\ref{fig:cavities}b, similar to those of Refs.~\cite{reagor2,pfaff}, but with a higher resonant frequency $\sim 9.4\,$GHz. The cooling field was applied normal to the bottom circular surface, which is therefore the only one that contributes to the trapped-flux dissipation; qualitatively similar results where obtained for fields perpendicular to the cavity axis.

The cavities were cooled down to 30~mK inside a mumetal can to shield them from ambient magnetic fields. Starting from temperature above $T_c$, different cooling fields were generated by Helmholtz coils placed inside the can and the fields were calibrated by measuring with a fluxgate magnetometer at room temperature for different applied coil currents. To measure cavity performance with different trapped fields, the refrigerator was warmed to approximately 1.5~K, and a current through the coil was applied until the samples cooled to millikelvin temperatures. We observed no significant change in cavity dissipation when either the fields were applied when well below $T_c$, nor when the magnet was turned off at low temperature.

\begin{figure}
  \centering
  \includegraphics[width=0.46\textwidth]{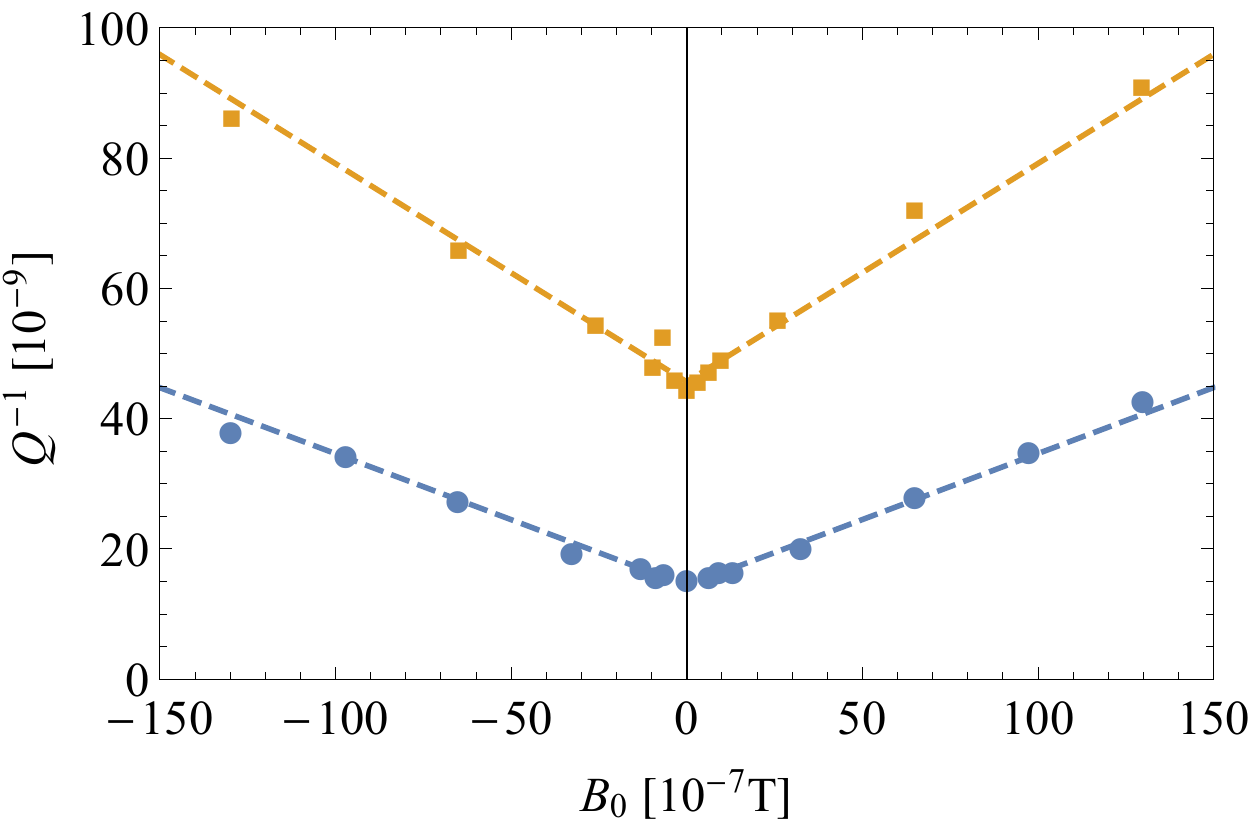}
  \caption{Inverse quality factor vs cooling field. Blue circles: measured quality factor for the coaxial cavity; orange squares: same for the rectangular cavity. Dashed lines are best linear fits to the experimental data.}\label{fig:expdata}
\end{figure}

We show in Fig.~\ref{fig:expdata} the results of the measurements for the coaxial cavity (circles) and the rectangular one (squares). The lines are best linear fits with intercepts corresponding to zero-field quality factors $Q_c=7\times 10^7$ and $Q_r=2.2\times 10^7$, and experimental slopes $\alpha_{c,e} = 0.0020\,$T$^{-1}$ and $\alpha_{r,e} = 0.0034\,$T$^{-1}$ for coaxial and rectangular cavity, respectively. 
The data display linear behavior down to less than 5~mG, indicating that the ambient field is smaller than that.
Both slopes agree with an order-of-magnitude estimate from Eq.~(\ref{eq:a_factors}), as we now show: first,
the critical field of aluminum is $B_c \simeq 0.01\,$T~\cite{AlHc}. The skin depth at low temperature can be estimated from the residual resistivity, which for 4N aluminum is of order $\rho_n \sim 10^{-10}\,\Omega\cdot$m~\cite{Ribot}; substituting this value and $\omega/2\pi= 10\,$GHz into Eq.~(\ref{eq:skindepth}) gives $\delta_s\sim 50\,$nm. However, this value is small compared to the mean free path $\ell = 3/\rho_n \nu e^2 v_F\sim 4\,\mu$m, where we used $\nu = 1.5\times10^{47}/$Jm$^3$ and $v_F = 2\times 10^6\,$m/s.
Therefore, we use instead Eq.~(\ref{eq:anom_skin}) to find $\delta_{s,a} \simeq 0.2\,\mu$m. Finally, the geometry factor $G$ is, for low-lying modes, of order of the inverse of the cavity size; the latter is $\sim c/\omega \sim 1\,$cm. Substituting these values into Eq.~(\ref{eq:a_factors}) we get $\alpha \sim 10^{-3}\,$T$^{-1}$, which agrees in order of magnitude with the values extracted from the experiments.

For a more accurate comparison between theory and experiment, we now improve our estimates for the geometry factor. 
To that end, we find it convenient to introduce a dimensionless geometry factor $\tilde{G}$ defined as
\begin{equation}\label{eq:dimG}
  \tilde{G} = \frac{\pi c}{2\omega}G
\end{equation}
For the coaxial cavity we find $\tilde{G}_c=2$, and for the rectangular cavity $\tilde{G}_r=1/b\sqrt{1/a^2+1/d^2}$, where $b$ is length in the electric field direction, $a$ and $d$ in the perpendicular directions, see Appendix~\ref{app:geometry}. In experiment we have $b \approx5\,$mm, $a\approx 17.8\,$mm, and $d\approx 31.3\,$mm; 
therefore we estimate $\tilde{G}_r \simeq 3.1$. From the dimensionless $\tilde{G}$ we obtain $G^{-1}_c \sim 0.40\,$cm and $G^{-1}_r \sim 0.25\,$cm, implying $\alpha_{c,t} \sim 0.0025\,$T$^{-1}$ and $\alpha_{r,t} \sim 0.0040\,$T$^{-1}$. The agreement of the two estimates with the respective experimental results is fairly good, given that the anomalous skin depth, Eq.~(\ref{eq:anom_skin}), is defined only up to a numerical factor of order unity.

\section{Summary and discussion}
\label{sec:summary}

When cooled in the presence of a magnetic field, superconductors can trap flux in the form of vortices in type II superconductors, or normal domains (tubes, laminae) in a type I material. The motion of these normal regions is responsible for dc dissipation. At low temperatures (when the quasiparticle density is exponentially suppressed), and at higher frequencies rendering pinning ineffective,  the dominant contribution to the ac absorption may come from two mechanisms: deformation of the superconducting/normal state interfaces and direct absorption at the surfaces of the normal regions exposed to ac fields. In this work, we have reviewed and extended the analysis of these mechanisms to include both type I and type II superconductors; in particular, for type I we study the deformation of the $S$/$N$ interface in Sec.~\ref{sec:t1def}. We find that the deformation of vortex lines is the dominant effect in type II superconductors, while the direct absorption is dominant in type I superconductors, as discussed in Sec.~\ref{sec:Pkappa}. 

We consider the dependence of the quality factor of a superconducting cavity on the cooling field in Sec.~\ref{sec:Qfactor}. Focusing on type I superconductors, we present experimental data for aluminum cavities of two shapes, rectangular and coaxial. The measured reductions of the quality factors in increasing cooling field are in good agreement with theoretical calculations based on independent estimates of material parameters. With our findings, we can answer an important question for applications: what is the maximum field $B_0^\mathrm{max}$ in which one can cool a cavity while maintaining a high quality factor? As an example, let us assume that $Q\sim 10^{9}$ is targeted in an aluminum cavity; then inverting Eq.~(\ref{eq:QiB0}) and using $\alpha \sim 3\times 10^{-3}$T$^{-1}$, we estimate $B_0^\mathrm{max} \sim 3\times 10^{-7}$T (\textit{i.e.}, a few milliGauss). Since with careful shielding fields smaller than this (about 1~mG) can be obtained, dissipation due to the trapped flux does not necessarily limit the quality factor of such Al cavities to the measured $\lesssim 10^9$ values ~\cite{reagor,kudra}. In fact, taken together, our estimate for $B_0^\mathrm{max}$, the observed linear behavior down to few mG, and the finite intercept corresponding to $Q < 10^{8}$ in Fig.~\ref{fig:expdata}, indicate that our cavities are limited by some other, independent of field dissipation mechanism (such as dielectric losses, two-level systems, or non-equilibrium quasiparticles). Therefore, further improvements in magnetic shielding to reduce the cooling field are not likely to improve the cavity quality factors, even though at 1~mG the flux trapped \textit{e.g.} at the bottom of the coaxial cavity corresponds to over $10^4$ magnetic flux quanta.

Recently, niobium cavities designed for particle accelerators have been considered also for quantum information applications~\cite{romanenko1}. Given the room temperature resistivity ($1.5\times 10^{-7}\Omega\cdot$m) and residual resistivity ratio ($\sim 200$), using Eq.~(\ref{eq:skindepth}) we estimate $\delta_s \sim 4\times 10^{-7}\,$m at 1.3~GHz. With $B_c\sim 0.2\,$T~\cite{Liarte2017} and $G \sim \omega/c\sim 30\,$m$^{-1}$, from Eq.~(\ref{eq:a_factors}) we find $\alpha_\mathrm{Nb} \sim 3\times 10^{-5}$T$^{-1}$, indicating that Nb cavities are much less affected by ambient field than Al ones. We note that using type-I formulas is adequate for an order-of-magnitude estimate, since $\kappa\sim0.73\,\text{--}\,1.5$ for Nb~\cite{Liarte2017,ooi}; in fact, cavity-grade Nb behaves as a type-II/1 superconductor in which, due to attraction at long distances, vortices form bundles interspersed by Meissner state regions, a state known as intermediate mixed state~\cite{ooi}. Interestingly, cooling protocols under which flux can be expelled have been developed~\cite{romanenko2}. Without flux expulsion, the quality factor in a cooling field $B_0= 10^{-6}\,$T was measured to be about $1.5\times10^{10}$, in reasonable agreement with the estimate $Q=1/\alpha_\mathrm{Nb}B_0 \sim 3\times10^{10}$. Moreover, measurements of the temperature dependence of the quality factors, both before and after heat treatments, confirm the limiting effect of two-level systems in Nb cavities with low-temperature $Q\lesssim 2\times10^{10}$~\cite{romanenko1}, in qualitative agreement with our analysis for Al cavities. While these results point to the need to further improvements in the material properties and surface treatments for both Nb and Al, our findings establish that, even for Nb, careful shielding (or flux expulsion) during cooling is necessary to achieve record quality factors.

\acknowledgments

We acknowledge useful discussions with Alex Gurevich.
The experimental work was supported by ARO grant W911NF-18-1-0212, and the theory effort by DOE contract DE-FG02-08ER46482
(L.I.G.), and by Feodor Lynen Research Fellowship and ARO grant W911NF-18-1-0212 (G.C.). 
The views and conclusions contained in this document are those of the authors and should not be interpreted as representing the official policies, either expressed or implied, of the ARO, the DOE or the U.S. Government. The U.S. Government is authorized to reproduce and distribute reprints for Government purposes notwithstanding any copyright notation herein.
L.F. and R.J.S. are founders and shareholders of Quantum Circuits, Inc.

\appendix

\section{Vortex core contribution to line tension}
\label{app:coretension}

Here we consider in more detail the line tension $\varepsilon$ for vortices in type II superconductors. In Sec.~5.1.2 of Ref.~\cite{Tinkham} the formula in Eq.~(\ref{eq:tensionII}) is obtained by considering the energy of currents and fields \textit{outside} the vortex core in the limit $\kappa \gg 1$. In principle, there is energy associated with the bending of the core, which is neglected there. We now substantiate such approximation for type II superconductors. 

We start by considering a flux tube in a type I superconductor; we indicate below how to extend the final result to the bending of a vortex core.
We treat the flux tube as a cylinder of radius $R_t$ and we denote with $\gamma$ the surface energy associated with a domain wall separating a normal region from a superconducting one. Consider a small piece of height $dz$ of the flux tube: when the top of the small piece is displaced perpendicularly to the $z$ direction by a small amount $du$, the change in energy due to deformation of the surface is
\begin{align}
    \Delta E & = \gamma \Delta A \simeq \gamma 2\pi R_t \left(\sqrt{(dz)^2+ (du)^2} - dz \right) \nonumber \\ & \simeq \gamma 2\pi R_t dz \frac12 \left(\frac{du}{dz}\right)^2\,,
\end{align}
From this expression, we estimate the bending contribution $\varepsilon_b$ to the line tension to be
\begin{equation}
    \varepsilon_b = 2\pi \gamma R_t\,.
    \label{eq:Rt}
\end{equation}

As discussed in Sec.~\ref{sec:t1def}, the surface energy is [cf. Eq.~(\ref{eq:surfen})]
\begin{equation}
    \gamma = \delta \frac{B_i^2}{2\mu_0}
\end{equation}
with $\delta \sim \xi -\lambda$ and $B_i$ the field in the normal region. The latter is the critical field $B_c$ in a type I material, but may in general differ from it. In fact, we may apply Eq.~(\ref{eq:Rt}) to a type II superconductor by setting $R_t\approx\xi$ and using for $H_i$ the vortex core field [see Eq.~(5.14b) in~\cite{Tinkham}],
\begin{equation}
    H_i = \frac{\Phi_0}{2\pi\lambda^2} \ln \kappa\,.
\end{equation}
This yields
\begin{equation}
    \varepsilon_b \simeq -2\pi \xi \lambda\frac{\Phi_0^2}{4\pi^2\mu_0 \lambda^4} \ln^2 \kappa = -2\frac{\ln \kappa}{\kappa} \varepsilon
\end{equation}
where in the last formula we use $\varepsilon$ of Eq.~(\ref{eq:tensionII}) [that is, the main contribution to the vortex line tension originating from outside the core]. Since $\kappa \gg 1$, we find $|\varepsilon_b| \ll \varepsilon$.

\section{Dimensionless geometry factor}
\label{app:geometry}

We sketch here the calculation of the dimensionless geometry factor $\tilde{G}$ for the two cases of interest, a coaxial cavity and a rectangular one, see Fig.~\ref{fig:cavities}.

For the coaxial cavity, the magnetic field for the TEM mode in a $\lambda/4$ resonator can be written in cylindrical coordinates $\{\rho,\theta,z\}$ as~\cite{pozar}
\begin{equation}
    \bar{H}(\rho,z) = \hat{\theta} H_m \frac{a}{\rho}\frac{1}{\log a/b}\cos\frac{\pi z}{2L}
\end{equation}
where $H_m$ is the maximum value of the magnetic field in the cavity, $a$ is the radius of the inner conductor, $b$ is that of the outer one, and $L=\lambda/4$ is the length of the inner conductor. The shorted part of the coaxial cavity is at $z=0$ and the open part at $z=L$. The resonant frequency is $\omega = 2\pi c/\lambda =\pi c/2L$. Integrations of $H^2(\rho,z)$ over the volume of the cavity and of $H^2(\rho,0)$ over the bottom surface are straightforward and, using the definitions in Eqs.~(\ref{eq:Gdef}) and (\ref{eq:dimG}), give $\tilde{G}_c=2$.

For the TE$_{101}$ mode of a rectangular cavity, if the electric field is pointing in the $y$ direction, the magnetic field has components in the $x$ and $z$ directions~\cite{pozar}. We don't need to know the spatial profile of the field:
the cavity volume is in the region $0<x<a$, $0<y<b$, $0<z<d$ and the surfaces with trapped flux are those at $y=0$ and $y=b$. Then from Eq.~(\ref{eq:Gdef}) one immediately finds $G_r=2/b$, since the integrals over variables $x$ and $z$ are the same in the numerator and in the denominator. The result for $\tilde{G}_r$ follows from $\omega = c\sqrt{(\pi/a)^2+(\pi/d)^2}$~\cite{pozar}.


\begin{thebibliography}{60}
\bibitem{note2}In the case of thin-film devices in the presence of an out-of-plane magnetic field, one can generically expect type II behavior: thin films are usually disordered superconductors with $\ell \ll \xi_0$, where $\xi_0$ is the coherence length of the clean material. For disordered superconductors the Ginzburg-Landau parameter is approximately given by $\kappa \approx \lambda/\ell$ (see Sec.~4.2.1 in~\cite{Tinkham}), which is larger than the corresponding ``clean'' value. Moreover, for thickness $d$ small compared to $\lambda$, the length scale $\lambda_\perp$ that determines the extent of the supercurrent surrounding the vortex core is $\lambda_\perp \approx \lambda^2/d$ (see Sec.~3.11.4 in~\cite{Tinkham}); this length scale is longer than the bulk value $\lambda$, so that using $\lambda_\perp$ to calculate $\kappa$ leads to a further increase in its value compared to that for the bulk material.
\bibitem{note4}To avoid confusion, we note that the geometry factor introduced here is not the same used in the accelerator community, $G_\mathrm{acc}$. However, if all surfaces contribute to loss by trapped flux, they are related as $G_\mathrm{acc} = \pi Z_0/2\tilde{G}$, where $Z_0$ is the vacuum impedance and the dimensionless factor $\tilde{G}$ is defined in Eq.~(\ref{eq:dimG}).
\end{thebibliography}
\end{document}